# *Single beam optical conveyor belt for chiral particles*

*David E. Fernandes*[1*], *Mário G. Silveirinha*[1,2†]

[1]*Department of Electrical Engineering, University of Coimbra, and Instituto de Telecomunicações, 3030-290 Coimbra, Portugal*

[2]*University of Lisbon, Instituto Superior Técnico, Avenida Rovisco Pais, 1, 1049-001 Lisboa, Portugal*

**Abstract**

A different paradigm is proposed to selectively manipulate and transport small engineered chiral particles and discriminate different enantiomers using unstructured chiral light. It is theoretically shown that the response of a chiral metamaterial particle may be tailored to enable an optical conveyor belt operation with no optical traps, such that for a fixed incident light helicity the nanoparticle is either steadily pushed towards the direction of the photon flow or steadily pulled against the photon flow, independent of its position. Our findings create distinct opportunities for unconventional optical manipulations of tailored nanoparticles and may have applications in sorting racemic mixtures of artificial chiral molecules and in particle delivery.

[*] E-mail: dfernandes@co.it.pt

[†] To whom correspondence should be addressed: E-mail: mario.silveirinha@co.it.pt

# I. Introduction

The optical manipulation and trapping of micro- and nano-sized particles or atoms is an important topic of research [1-3] with applications in laser cooling of neutral atoms, trapping of single cells, biosensors, particle transport, and in microfluidic cell sorting [4, 5]. These manipulations usually rely on tightly focused laser beams (optical tweezers [6]) and in field intensity gradients, which allow to precisely position and move neutral particles or atoms [7, 8]. In recent years, unconventional light-matter interactions relying on unstructured light (i.e. with no gradient along the direction of propagation) have received a great deal of attention. Conventional wisdom suggests that when a polarizable particle is illuminated by a light beam the momentum transfer should always push the particle downstream, i.e., along the direction of the light flow. Remarkably, different works shattered this understanding and demonstrated the counterintuitive opportunity of having optomechanical interactions that induce pulling forces which drag a particle in a direction opposite to the light flow [9-15]. Usually this is done by engineering the form in which the particle scatters light in such manner that the forward scattering is maximized, so that the wave momentum is effectively reinforced in the forward direction [12-15]. To compensate for the increased light momentum in the forward direction, the object experiences a pulling force that drags it towards the light source. A diffractionless light beam with the ability to exert a negative radiation pressure is typically referred to as “tractor beam”.

Other solutions to obtain negative radiation pressures and related phenomena have also been proposed, both theoretically and experimentally [16-26]. For instance, it was demonstrated – relying either on the interference of two co-propagating or counter-propagating waves – that it is possible to sort, organize or have a bi-directional transport of sub-micron particles [16-18]. In particular, it was shown that by actively controlling the relative phase between two co-propagating Bessel beams it is feasible to have an optical conveyor belt operation such that micron-sized particles can be transported both in the upstream and downstream directions [18].

Here, inspired by our previous work [27], we propose a different paradigm for an optical conveyor belt that does not require a complex dynamic relative phase control of two beams as in Ref. [18]. We theoretically prove that by using a single light beam (e.g. a plane wave) and an opaque mirror, it is possible to have either a persistent pulling force or a persistent pushing force acting on an engineered chiral particle, such that the force sign is independent of the particle position with respect to the light source. Moreover, the sign of the optical force can be controlled simply by switching the helicity of the incident light. These findings generalize the theory of Ref. [27], wherein an analogous optical conveyor belt regime was demonstrated for a chiral slab infinitely extended along the directions parallel to the mirror. Here, it is proven that the same working principles can be exploited even for small chiral particles wherein the diffraction and scattering mechanisms are more complex. It should be noted that the emergence of negative pulling forces in chiral particles has been previously discussed in the literature relying on Bessel beams [25]. Moreover, it has also been proven both theoretically and experimentally that chiral microresonators may experience optical forces and torques strongly dependent on the wave polarization [28]. Recently, it was also shown that chiral particles standing on the top a dielectric substrate can experience an anomalous lateral force [26].

This article is organized as follows. In Section II a theoretical formalism is developed to characterize the optical force when a chiral particle placed in front of an opaque mirror is illuminated by unstructured chiral light. The chiral particle is modeled as a superposition of electric and magnetic dipoles. It is theoretically demonstrated that by tailoring the chiral and magnetic responses it is possible to have an optical conveyor belt operation with no optical traps. To illustrate the proposed effect, in Section III we present a numerical study of the optical force exerted on a conjugated gammadions chiral particle. Finally, in Section IV the conclusions are drawn. The time dependence of the electromagnetic fields is assumed of the type $e^{-i\omega t}$.

## II. Passive Optical Conveyor Belt

The idea to realize a passive optical conveyor belt is to illuminate an engineered chiral particle in front of an opaque mirror with unstructured chiral light [27] (Fig. 1). The chiral particle must have a strongly asymmetric response such that it is approximately transparent to light with a specific helicity (let us say, left circularly polarized (LCP) light), whereas it should strongly absorb light with the opposite helicity (let us say, right circularly polarized (RCP) light). Under these conditions, one may expect that if the incident wave is RCP then the particle will be able to extract a significant momentum from the incident beam and hence it will be pushed downstream towards the mirror. On the other hand, if the incident wave is LCP it will interact weakly with the particle, as it is approximately transparent for this polarization. Interestingly, upon reflection on the mirror the wave polarization is switched from LCP to RCP. Thus, the wave reflected on the mirror interacts strongly with the chiral particle and hence will impart a negative momentum on it, originating in this way a pulling (upstream) force [27]. Therefore, depending on the incident wave polarization the chiral particle is either steadily pushed downstream (positive force) or upstream (negative force). Note that this optical conveyor belt operation is passive in the sense that for a fixed direction of particle delivery the incident wave properties do not require any type of control. To investigate under which conditions the proposed intuitive picture may have correspondence with the real electromagnetic situation, next we rigorously determine the optical force acting on a chiral particle modeled as a point dipole.

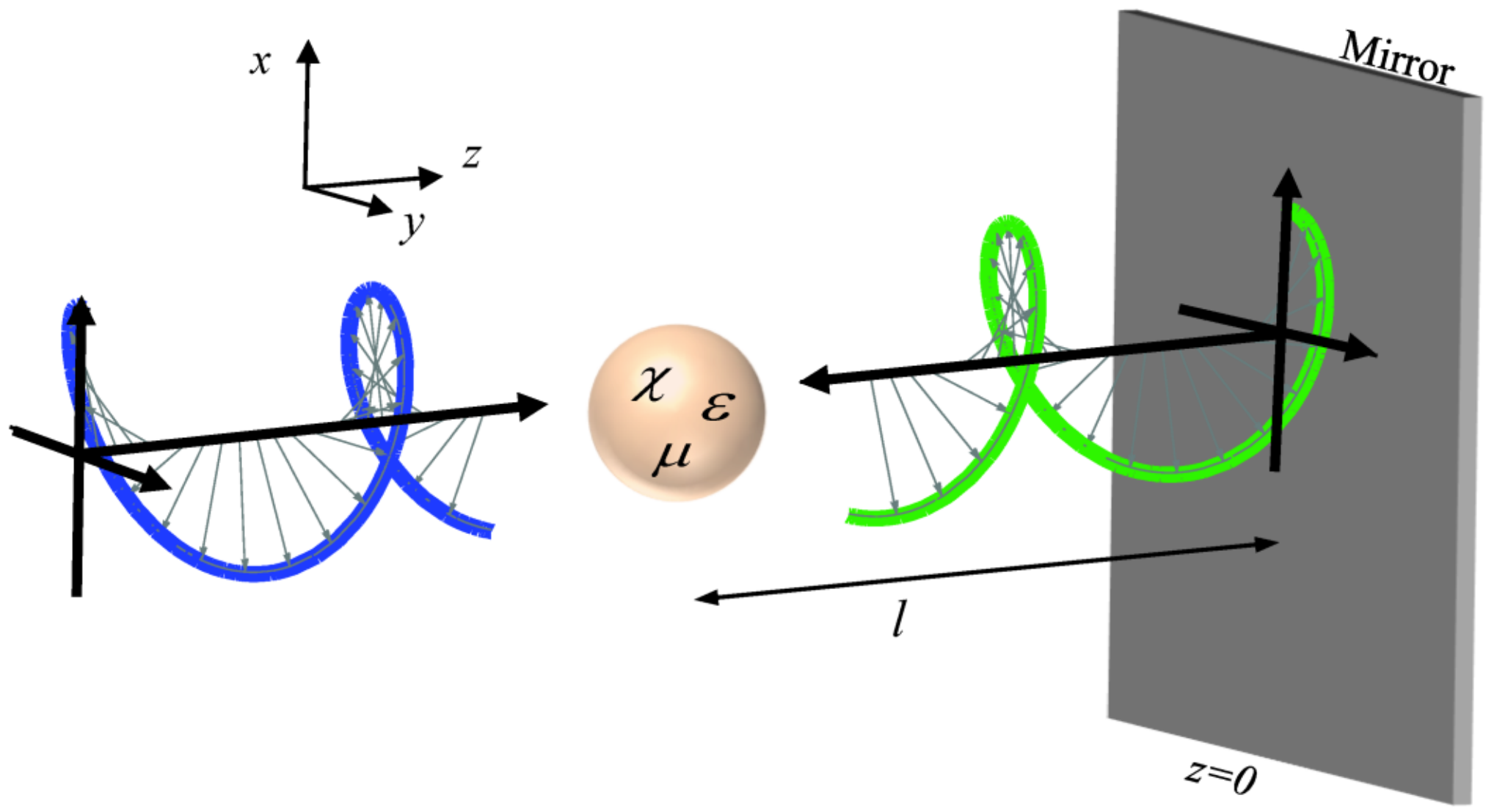

Fig. 1 A chiral particle standing in free-space at a distance $l$ from an opaque mirror is illuminated along the normal direction by an incident circularly polarized plane wave.

### A. *The optical force*

The optical force exerted on electrically small particles standing in free-space depends on the *local* electromagnetic field, i.e. the field that effectively polarizes the particle, and on the electric and magnetic dipole moments, $\mathbf{p}_e$ and $\mathbf{p}_m$ respectively. We define $\mathbf{p}_m$ as $\mathbf{p}_m = \mu_0 \mathbf{m}$, where $\mathbf{m}$ is the standard textbook magnetic dipole moment with units of $\mathrm{A} \times \mathrm{m}^2$. The $i$-th component of the time-averaged optical force can be written in a compact manner as [12, 29-31],

$$\mathscr{F}_i = \frac{\varepsilon_0}{2} \mathrm{Re}\left\{ \mathbf{P}^* \cdot \partial_i \mathbf{F}_{\mathrm{loc}} - \frac{1}{6\pi} \frac{\omega^4}{c^2} \eta_0 \left( \mathbf{p}_e \times \mathbf{p}_m^* \right) \cdot \hat{\mathbf{u}}_i \right\}, \qquad (i=x,y,z) \qquad (1)$$

where $\hat{\mathbf{u}}_i$ is a unit vector along the $i$-th direction, $\partial_i \equiv \hat{\mathbf{u}}_i \cdot \nabla$, $\varepsilon_0$ is the vacuum permittivity, $\mu_0$ is the free-space permeability, and $\eta_0$ is the free-space impedance. In the above, the six component vector $\mathbf{F}_{\text{loc}} = \begin{pmatrix} \mathbf{E}_{\text{loc}} \\ \mathbf{H}_{\text{loc}}\eta_0 \end{pmatrix}$ represents the local electromagnetic fields that excite the particle, and $\mathbf{P} = \begin{pmatrix} \mathbf{p}_e \varepsilon_0^{-1} \\ \mathbf{p}_m c \end{pmatrix}$ is a six-component vector with the dipole moment components. The second term in Eq. (1) is the recoil force due to the self-interaction of the electric and magnetic dipoles which was derived in Ref. [30]. In the examples considered in this article we found that this second term is negligible.

The dipole moments $\mathbf{P}$ excited in the chiral particle are related to the local field $\mathbf{F}_{\text{loc}}$ as follows:

$$\mathbf{P} = \underbrace{\begin{pmatrix} \overline{\overline{\alpha}}_{ee} & \overline{\overline{\alpha}}_{em} \\ \overline{\overline{\alpha}}_{me} & \overline{\overline{\alpha}}_{mm} \end{pmatrix}}_{\overline{\overline{\alpha}}} \cdot \mathbf{F}_{\text{loc}}, \tag{2}$$

where $\overline{\overline{\alpha}}$ is a 6×6 matrix with sub-components $\overline{\overline{\alpha}}_{ee}$, $\overline{\overline{\alpha}}_{em}$, $\overline{\overline{\alpha}}_{me}$, $\overline{\overline{\alpha}}_{mm}$, which are the 3×3 polarizability tensors of the particle and have dimensions of volume $\left[\text{m}^3\right]$. The crossed polarizabilities $\overline{\overline{\alpha}}_{em}$ and $\overline{\overline{\alpha}}_{me}$ determine the magneto-electric response and may be non-zero for particles without inversion symmetry. For example, the magneto-electric response is non-zero for particles that cannot be superimposed on their mirror images, e.g., chiral particles [32]. Because chiral particles have an intrinsic handedness their electromagnetic response depends on the wave helicity. As a consequence, chiral-type structures enable phenomena such as polarization conversion [33], polarization rotation (optical activity) [34-36], polarization-dependent negative refraction [37-40] and a strongly asymmetric transmission of electromagnetic waves [27, 41-44]. It is well-known that for reciprocal media $\overline{\overline{\alpha}}_{em}$ and $\overline{\overline{\alpha}}_{me}$ are required to satisfy $\overline{\overline{\alpha}}_{me} = -\overline{\overline{\alpha}}_{em}^{T}$, where the superscript "*T*" denotes the operation of matrix transposition [32].

The incident wave radiated by the light source is assumed to be a circularly polarized plane wave propagating along the $+z$ direction, so that the incident field satisfies:

$$\mathbf{F}_{\mathrm{inc}}(\mathbf{r})=\begin{pmatrix}\mathbf{E}_{\mathrm{inc}}(\mathbf{r})\\ \eta_0\mathbf{H}_{\mathrm{inc}}(\mathbf{r})\end{pmatrix}=\begin{pmatrix}\mathbf{e}_{\pm}\\ -(\pm i)\mathbf{e}_{\pm}\end{pmatrix}E^{inc}e^{ik_0 z}, \qquad (3)$$

where $E^{inc}$ is the complex amplitude of the incident field, $\mathbf{e}_{\pm}=(\hat{\mathbf{x}}\pm i\hat{\mathbf{y}})/\sqrt{2}$, and $k_0=\omega/c$ is the free-space wave number. The subscript $\pm$ determines the polarization state of the incident wave, so that it is either RCP (+) or LCP (–). Evidently, without the opaque mirror the local field is coincident with the incident field: $\mathbf{F}_{\mathrm{loc}}=\mathbf{F}_{\mathrm{inc}}$. However, in the presence of the mirror (at $z=0$) the local field must also include the total field scattered by the mirror, which can be conveniently decomposed as $\mathbf{F}_{\mathrm{scat}}=\mathbf{F}_{\mathrm{ref}}+\mathbf{F}_{\mathrm{s,dip}}$, where $\mathbf{F}_{\mathrm{ref}}$ and $\mathbf{F}_{\mathrm{s,dip}}$ represent the waves generated upon reflection of the incident plane wave and of the field radiated by the point particle, respectively. Hence, we can write:

$$\mathbf{F}_{\mathrm{loc}}=\mathbf{F}_{\mathrm{inc}}+\mathbf{F}_{\mathrm{ref}}+\mathbf{F}_{\mathrm{s,dip}}. \qquad (4)$$

Without loss of generality, it is assumed that the mirror can be approximated by a perfect electric conductor (PEC). Therefore, $\mathbf{F}_{\mathrm{ref}}$ is given by:

$$\mathbf{F}_{\mathrm{ref}}(\mathbf{r})=\begin{pmatrix}\mathbf{E}_{\mathrm{ref}}(\mathbf{r})\\ \eta_0\mathbf{H}_{\mathrm{ref}}(\mathbf{r})\end{pmatrix}=-\begin{pmatrix}\mathbf{e}_{\pm}\\ (\pm i)\mathbf{e}_{\pm}\end{pmatrix}E^{inc}e^{-ik_0 z}. \qquad (5)$$

The calculation of the wave $\mathbf{F}_{\mathrm{s,dip}}$ is more elaborate, and is discussed in the next sub-section.

### *B. The field created by the point-particle*

The field created by the chiral particle is the superposition of the wave radiated by the particle in free-space ($\mathbf{F}_{\mathrm{dip}}$) and the corresponding wave scattered by the opaque mirror ($\mathbf{F}_{\text{s,dip}}$). The field $\mathbf{F}_{\text{dip}} = \begin{pmatrix} \mathbf{E}_{\text{dip}} \\ \eta_0 \mathbf{H}_{\text{dip}} \end{pmatrix}$ is the wave radiated by the electric and magnetic dipole moments $\mathbf{p}_e$ and $\mathbf{p}_m$, respectively, and satisfies the Maxwell equations:

$$\nabla \times \mathbf{E}_{\text{dip}} = i\omega\mu_0 \mathbf{H}_{\text{dip}} - \mathbf{j}_m, \tag{6a}$$

$$\nabla \times \mathbf{H}_{\text{dip}} = -i\omega\varepsilon_0 \mathbf{E}_{\text{dip}} + \mathbf{j}_e, \tag{6b}$$

where $\mathbf{j}_e(\mathbf{r}) = -i\omega \mathbf{p}_e \delta(\mathbf{r} - \mathbf{r}_0)$ is the electric current density associated with the electric dipole, $\mathbf{j}_m(\mathbf{r}) = -i\omega \mathbf{p}_m \delta(\mathbf{r} - \mathbf{r}_0)$ is the magnetic current density associated with the magnetic dipole, and $\mathbf{r}_0 = (x_0, y_0, z_0)$ is the position of the point-particle. Using the superposition principle, the field $\mathbf{F}_{\text{dip}}$ can be written in terms of the free-space Green function $\Phi(\mathbf{r}, \mathbf{r}_0) = \dfrac{e^{ik_0|\mathbf{r}-\mathbf{r}_0|}}{4\pi|\mathbf{r}-\mathbf{r}_0|}$ as follows:

$$\mathbf{F}_{\text{dip}}(\mathbf{r}, \mathbf{r}_0) = \begin{pmatrix} \mathbf{G}^{ee}(\mathbf{r}, \mathbf{r}_0) & \mathbf{G}^{em}(\mathbf{r}, \mathbf{r}_0) \\ \mathbf{G}^{me}(\mathbf{r}, \mathbf{r}_0) & \mathbf{G}^{mm}(\mathbf{r}, \mathbf{r}_0) \end{pmatrix} \cdot \mathbf{P}, \tag{7}$$

with

$$\mathbf{G}^{ee}(\mathbf{r}, \mathbf{r}_0) = \mathbf{G}^{mm}(\mathbf{r}, \mathbf{r}_0) \equiv \left(\nabla\nabla + \mathbf{1}k_0^2\right)\Phi(\mathbf{r}, \mathbf{r}_0), \tag{8a}$$

$$\mathbf{G}^{me}(\mathbf{r}, \mathbf{r}_0) = -\mathbf{G}^{em}(\mathbf{r}, \mathbf{r}_0) = -ik_0 \nabla\Phi(\mathbf{r}, \mathbf{r}_0) \times \mathbf{1}, \tag{8b}$$

where $\mathbf{1}$ is the 3×3 identity dyadic.

The opaque mirror scatters the wave $\mathbf{F}_{\text{dip}}$ creating in the this manner a reflected wave $\mathbf{F}_{\text{s,dip}}$. It is well known that for a PEC mirror $\mathbf{F}_{\text{s,dip}}$ can be calculated using the image method [45]. Specifically, the field $\mathbf{F}_{\text{s,dip}}$ may be regarded as the field radiated in free-space by fictitious image dipoles $\mathbf{p}'_e, \mathbf{p}'_m$ placed at the image point $\mathbf{r}'_0 = (x_0, y_0, -z_0)$ (Fig. 2).

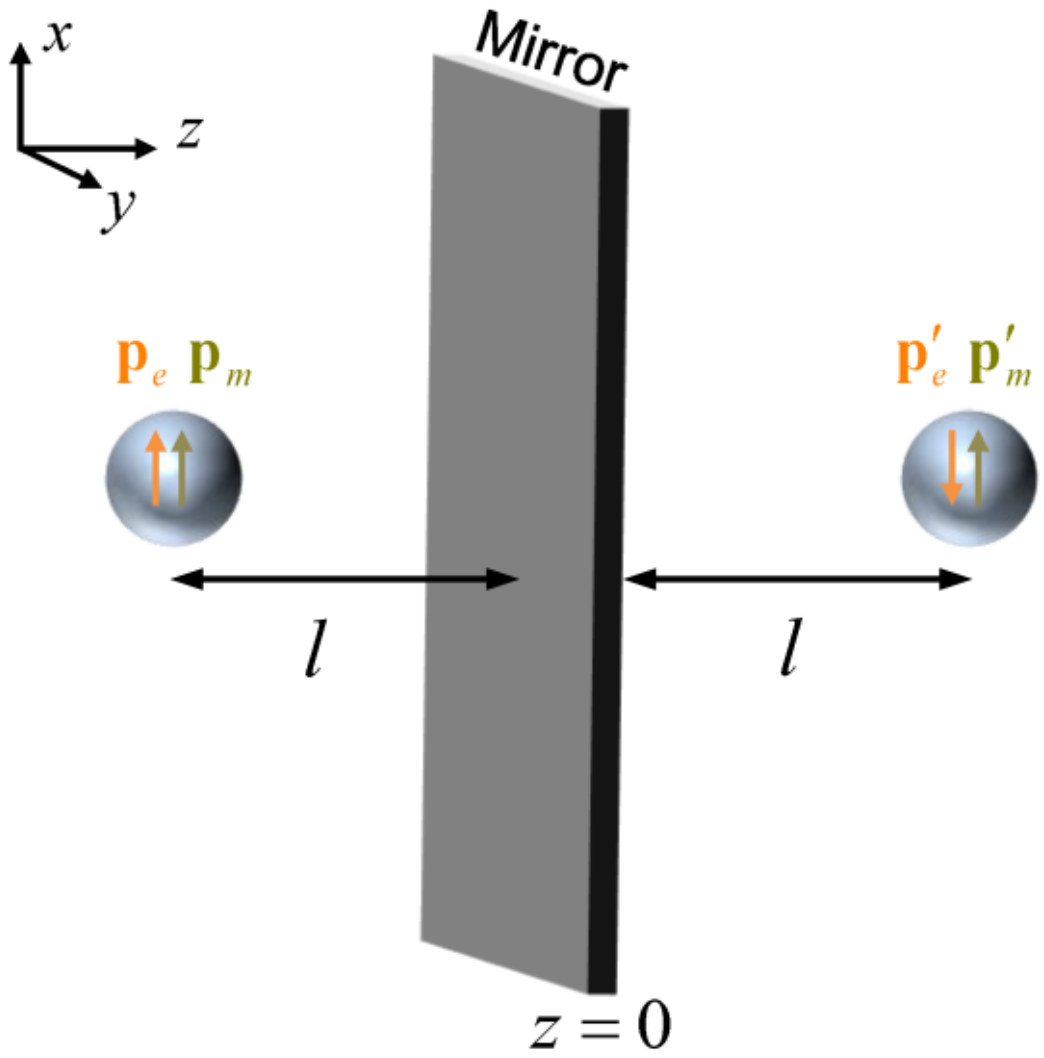

Fig. 2 Electric and magnetic dipoles $\mathbf{p}_e, \mathbf{p}_m$ with coordinates $\mathbf{r}_0 = (x_0, y_0, z_0)$ are at a distance $l = -z_0$ from a metallic mirror ( $z = 0$ ). From the image method, the mirror may be replaced by fictitious image dipoles $\mathbf{p}'_e, \mathbf{p}'_m$ with coordinates $\mathbf{r}'_0 = (x_0, y_0, -z_0)$ without affecting the electromagnetic field distribution.

For a PEC mirror the dipole moments of the fictitious image dipoles are related to the dipole moments of the point particle as [45]:

$$\mathbf{P}' \equiv \begin{pmatrix} \mathbf{p}'_e \varepsilon_0^{-1} \\ \mathbf{p}'_m c \end{pmatrix} = \begin{pmatrix} (-\mathbf{1}_t + \hat{\mathbf{z}}\hat{\mathbf{z}}) & \mathbf{0} \\ \mathbf{0} & (\mathbf{1}_t - \hat{\mathbf{z}}\hat{\mathbf{z}}) \end{pmatrix} \cdot \mathbf{P} \tag{9}$$

where $\mathbf{1}_t = \hat{\mathbf{x}}\hat{\mathbf{x}} + \hat{\mathbf{y}}\hat{\mathbf{y}}$ . Evidently, the field radiated by the image dipoles $\mathbf{F}_{\text{s,dip}}$ can be obtained from Eq. (7) by replacing $\mathbf{P} \to \mathbf{P}'$ and $\mathbf{r}_0 \to \mathbf{r}'_0$ , so that:

$$\mathbf{F}_{\text{s,dip}}(\mathbf{r}, \mathbf{r}'_0) = \begin{pmatrix} \mathbf{G}^{ee}(\mathbf{r}, \mathbf{r}'_0) & \mathbf{G}^{em}(\mathbf{r}, \mathbf{r}'_0) \\ \mathbf{G}^{me}(\mathbf{r}, \mathbf{r}'_0) & \mathbf{G}^{mm}(\mathbf{r}, \mathbf{r}'_0) \end{pmatrix} \cdot \mathbf{P}', \tag{10}$$

In particular, using Eq. (9) it follows that $\mathbf{F}_{\text{s,dip}}$ calculated at the position of the point-particle can be written as:

$$\mathbf{F}_{\text{s,dip}}(\mathbf{r}_0) = \mathbf{C}_{\text{int}}(\mathbf{r}_0, \mathbf{r}'_0) \cdot \mathbf{P}, \tag{11}$$

where

$$\mathbf{C}_{\text{int}}\left(\mathbf{r}_0,\mathbf{r}_0'\right)=\begin{pmatrix}\mathbf{G}^{ee}\left(\mathbf{r}_0,\mathbf{r}_0'\right) & \mathbf{G}^{em}\left(\mathbf{r}_0,\mathbf{r}_0'\right)\\ \mathbf{G}^{me}\left(\mathbf{r}_0,\mathbf{r}_0'\right) & \mathbf{G}^{mm}\left(\mathbf{r}_0,\mathbf{r}_0'\right)\end{pmatrix}\cdot\begin{pmatrix}\left(-\mathbf{1}_t+\hat{\mathbf{z}}\hat{\mathbf{z}}\right) & \mathbf{0}\\ \mathbf{0} & \left(\mathbf{1}_t-\hat{\mathbf{z}}\hat{\mathbf{z}}\right)\end{pmatrix}, \tag{12}$$

is by definition the interaction matrix.

### *C. The dipole moments*

At this point, we are ready to determine the dipole moments induced by the incident wave on the point-particle. From Eqs. (4) and (11), one sees that the local field can be written as

$$\mathbf{F}_{\text{loc}}\left(\mathbf{r}_0\right)=\mathbf{F}_{\text{inc}}\left(\mathbf{r}_0\right)+\mathbf{F}_{\text{ref}}\left(\mathbf{r}_0\right)+\mathbf{C}_{\text{int}}\left(\mathbf{r}_0,\mathbf{r}_0'\right)\cdot\mathbf{P}\,. \tag{13}$$

Then, from the constitutive relation (2) that determines the electromagnetic response of the dipole, it is found that:

$$\mathbf{P}=\mathbf{M}\cdot\left(\mathbf{F}_{\text{inc}}+\mathbf{F}_{\text{ref}}\right),\quad \text{with}\quad \mathbf{M}=\left(\overline{\overline{\alpha}}^{-1}-\mathbf{C}_{\text{int}}\left(\mathbf{r}_0,\mathbf{r}_0'\right)\right)^{-1}. \tag{14}$$

Hence, for a given position of the particle $\mathbf{r}_0$ it is possible to compute the induced dipole moments using the above equation, and the optically induced force using Eqs. (1) and (13). In the absence of the mirror the force is obtained by setting $\mathbf{F}_{\text{ref}}=0$ and $\mathbf{C}_{\text{int}}\left(\mathbf{r}_0,\mathbf{r}_0'\right)=0$.

### *D. Condition for the optical conveyor-belt operation*

When the dipole is sufficiently far away from the mirror, the contribution of the mirror dipoles to the local field is expected to be negligible, i.e. $\mathbf{C}_{\text{int}}\approx 0$. In this situation the local field can be approximated by $\mathbf{F}_{\text{loc}}\left(\mathbf{r}_0\right)\approx\mathbf{F}_{\text{inc}}\left(\mathbf{r}_0\right)+\mathbf{F}_{\text{ref}}\left(\mathbf{r}_0\right)$, which gives:

$$\mathbf{F}_{\text{loc}}\approx 2iE^{inc}\begin{pmatrix}\sin\left(k_0 z_0\right)\mathbf{e}_{\pm}\\ -\left(\pm\cos\left(k_0 z_0\right)\right)\mathbf{e}_{\pm}\end{pmatrix}. \tag{15}$$

Using $\mathbf{P}\approx\overline{\overline{\alpha}}\cdot\mathbf{F}_{\text{loc}}$ in Eq. (1) and neglecting the self-interaction between the electric and magnetic dipoles it is found that the optical force can be approximated by:

$$\mathcal{F}_z = 2k_0\varepsilon_0 \left|E^{inc}\right|^2 \operatorname{Re}\left\{\begin{pmatrix} \sin(k_0 z_0)\mathbf{e}_\pm \\ -(\pm\cos(k_0 z_0))\mathbf{e}_\pm \end{pmatrix} \cdot \begin{pmatrix} \overline{\overline{\alpha}}_{ee} & \overline{\overline{\alpha}}_{em} \\ \overline{\overline{\alpha}}_{me} & \overline{\overline{\alpha}}_{mm} \end{pmatrix}^T \cdot \begin{pmatrix} \cos(k_0 z_0)\mathbf{e}_\pm \\ (\pm\sin(k_0 z_0))\mathbf{e}_\pm \end{pmatrix}^* \right\} \quad (16)$$

Using the reciprocity constraints, it follows that for an isotropic chiral particle the force reduces to:

$$\mathcal{F}_z = 2k_0\varepsilon_0 \left|E^{inc}\right|^2 \operatorname{Re}\left\{\begin{pmatrix} \sin(k_0 z_0) \\ -(\pm\cos(k_0 z_0)) \end{pmatrix} \cdot \begin{pmatrix} \alpha_{ee} & -\alpha_{em} \\ +\alpha_{em} & \alpha_{mm} \end{pmatrix} \cdot \begin{pmatrix} \cos(k_0 z_0) \\ (\pm\sin(k_0 z_0)) \end{pmatrix} \right\}, \quad (17)$$

being the "+" sign ("–" sign) associated with an RCP (LCP) incident wave. Thus, after straightforward simplifications it is found that:

$$\frac{\mathcal{F}_{z,LCP}}{2k_0 W_{\text{av}}^{inc}} = \operatorname{Re}\left\{\alpha_{ee} - \alpha_{mm}\right\}\sin(2k_0 z_0) + 2\operatorname{Re}\left\{\alpha_{em}\right\}. \quad (18a)$$

$$\frac{\mathcal{F}_{z,RCP}}{2k_0 W_{\text{av}}^{inc}} = \operatorname{Re}\left\{\alpha_{ee} - \alpha_{mm}\right\}\sin(2k_0 z_0) - 2\operatorname{Re}\left\{\alpha_{em}\right\}. \quad (18b)$$

where $W_{\text{av}}^{inc} = \frac{\varepsilon_0}{2}\left|E^{inc}\right|^2$ is the time-averaged energy density associated with the incoming wave. It is interesting to note that the difference between the optical forces associated with the two circular polarization states is always independent of the particle position $\mathcal{F}_{z,RCP} - \mathcal{F}_{z,LCP} = 8k_0 W_{\text{av}}^{inc} \operatorname{Re}\left\{\alpha_{em}\right\}$. The derived formulas show that the optical force has two components. The first component ("gradient force") depends on the specific position ($z_0$) of the particle and is determined only by the *reactive* response of the particle. The second component ("scattering force") is completely independent of the particle position and depends only on the *absorption* emerging from the magneto-electric coupling ($\operatorname{Re}\left\{\alpha_{em}\right\}$). Remarkably, in the presence of the mirror the scattering force is completely independent of the loss terms $\operatorname{Im}\left\{\alpha_{ee}\right\}$ and $\operatorname{Im}\left\{\alpha_{mm}\right\}$ because the radiation pressure due to these terms has a different sign for waves travelling along the $+z$ and $-z$ directions. It can be shown that without the mirror the optical force is given by $\mathcal{F}_z = W_{av}^{inc} k_0 \left[\operatorname{Im}\left\{(\alpha_{ee} + \alpha_{mm})\right\} - (\pm 2)\operatorname{Re}\left\{\alpha_{em}\right\}\right]$ for an incident wave travelling along the $+z$ direction. As before, the $+(-)$ sign is associated with the RCP (LCP) polarization state. For an incident wave travelling along the $-z$ direction the force flips sign.

Importantly, Eq. (18) unveils a quite interesting possibility: if the reactive response of the chiral particle is designed to be balanced, such that $\mathrm{Re}\left\{\alpha_{ee}-\alpha_{mm}\right\}=0$, then the optical force becomes independent of the particle position and has opposite signs for RCP and LCP incident waves. Remarkably, the absorption of electromagnetic radiation and the opaque mirror are essential to harness the sign of the optical force. A semi-transparent mirror would create a less intense reflected wave and hence would lead to weaker negative optical forces. It should be mentioned that other authors have shown that balanced chiral particles standing in free-space (satisfying a set of conditions more restrictive than ours) can be transparent to an incident circularly polarized electromagnetic field [46, 47].

The optical conveyor belt operation is possible even when the chiral particle is unbalanced ($\mathrm{Re}\left\{\alpha_{ee}-\alpha_{mm}\right\}\neq 0$). Indeed, it is easy to check that the signs of $\mathcal{F}_{z,LCP}$ and $\mathcal{F}_{z,RCP}$ are independent of $z_0$ provided:

$$\left|\mathrm{Re}\left\{\alpha_{ee}-\alpha_{mm}\right\}\right|<2\left|\mathrm{Re}\left\{\alpha_{em}\right\}\right|. \tag{19}$$

## III. Numerical Example

Next, the theory of Sect. II is applied to characterize the optical force acting on an artificially engineered (metamaterial) chiral particle. The cuboid metamaterial particle under analysis is formed by two conjugated-gammadions, as sketched in Fig. 3. Even though the particle is not isotropic, it is expected to behave effectively as an isotropic chiral particle when the incoming wave propagates along the $z$-direction.

In our design, the gammadions are made of silver and are embedded in polyimide. The polyimide and the gammadions are encapsulated between two layers of silicon, as shown in Fig. 3. Silver was modeled using the experimental data available in the literature [48]. The particle dimensions were optimized so that the magneto-electric response is boosted near $\lambda_0 = 1.55\,\mu\text{m}$, where high power solid-state lasers are available. For this wavelength the dimensions of the chiral particle, $a_x = a_y = a = 202.4\text{nm}$ and $a_z = 189.6\,\text{nm}$, are deeply subwavelength so that the dipole approximation is justified.

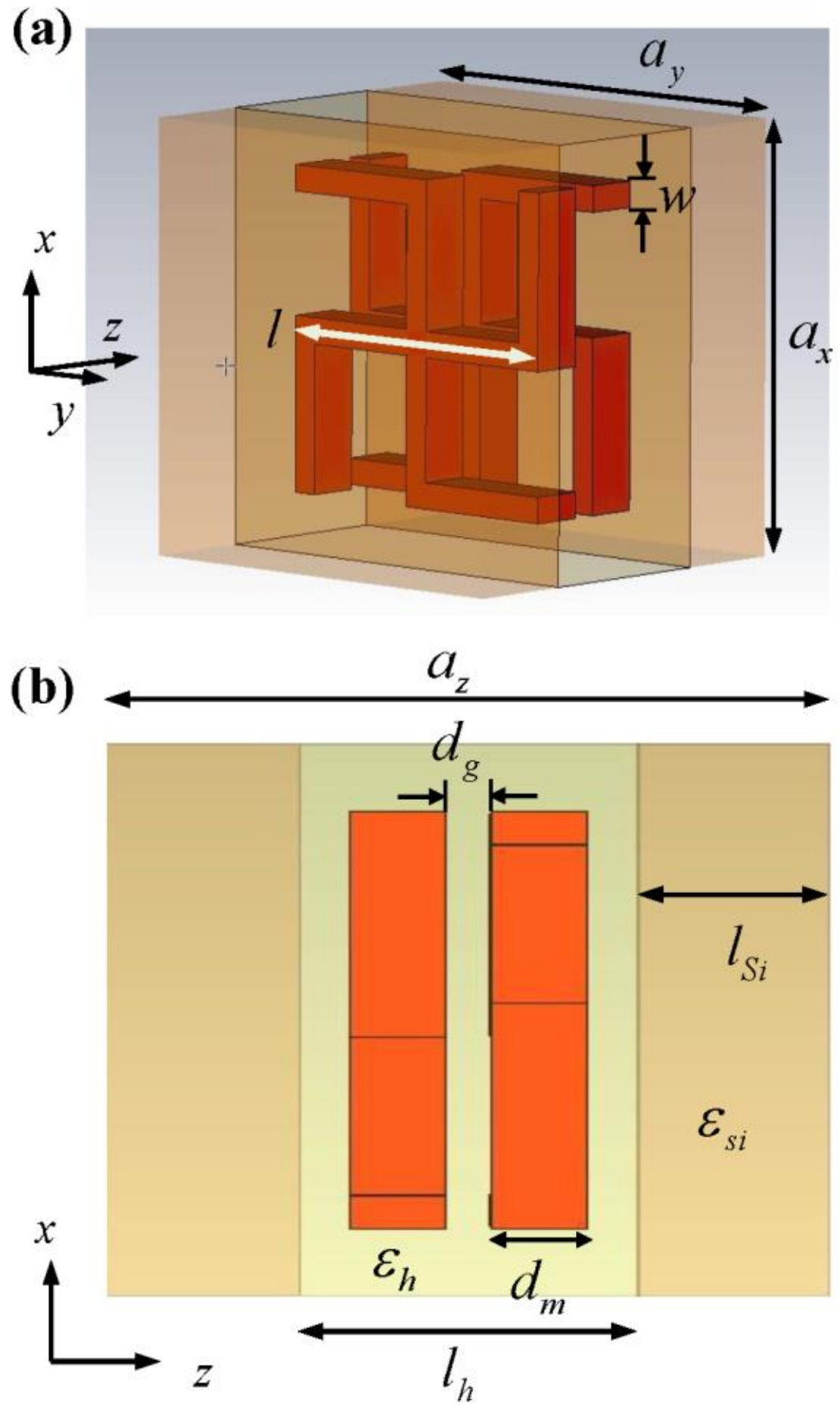

Fig. 3 **(a)** Geometry of the metamaterial particle. The length of the central arm of the gammadions is $l = 151.8\,\text{nm}$ and the width is $w = 12.2\,\text{nm}$. **(b)** Side view of the particle: the gammadions are made of silver with thickness $d_m = 25\,\text{nm}$ and are separated by a distance $d_g = 11.6\,\text{nm}$. The conjugated gammadions are embedded in a polyimide slab with thickness $l_h = 88.4\text{nm}$ and permittivity $\varepsilon_h = 6.25(1+i0.03)\varepsilon_0$ [35]. The polyimide slab is encapsulated between two layers of silicon, with permittivity $\varepsilon_{si} = 11.9(1+i0.004)\varepsilon_0$ and thickness $l_{si} = 50.6\,\text{nm}$. The total thickness of the cuboid along the $z$-direction is $a_z = 189.6\,\text{nm}$, and along the $x$- and $y$-directions is $a_x = a_y = a = 202.4\,\text{nm}$.

The polarizability matrix of the conjugated gammadions cuboid was calculated using the approach described in Ref. [49]. The basic idea is to excite the cuboid with a set of linearly polarized incident plane of waves and analyze the scattered far-field.

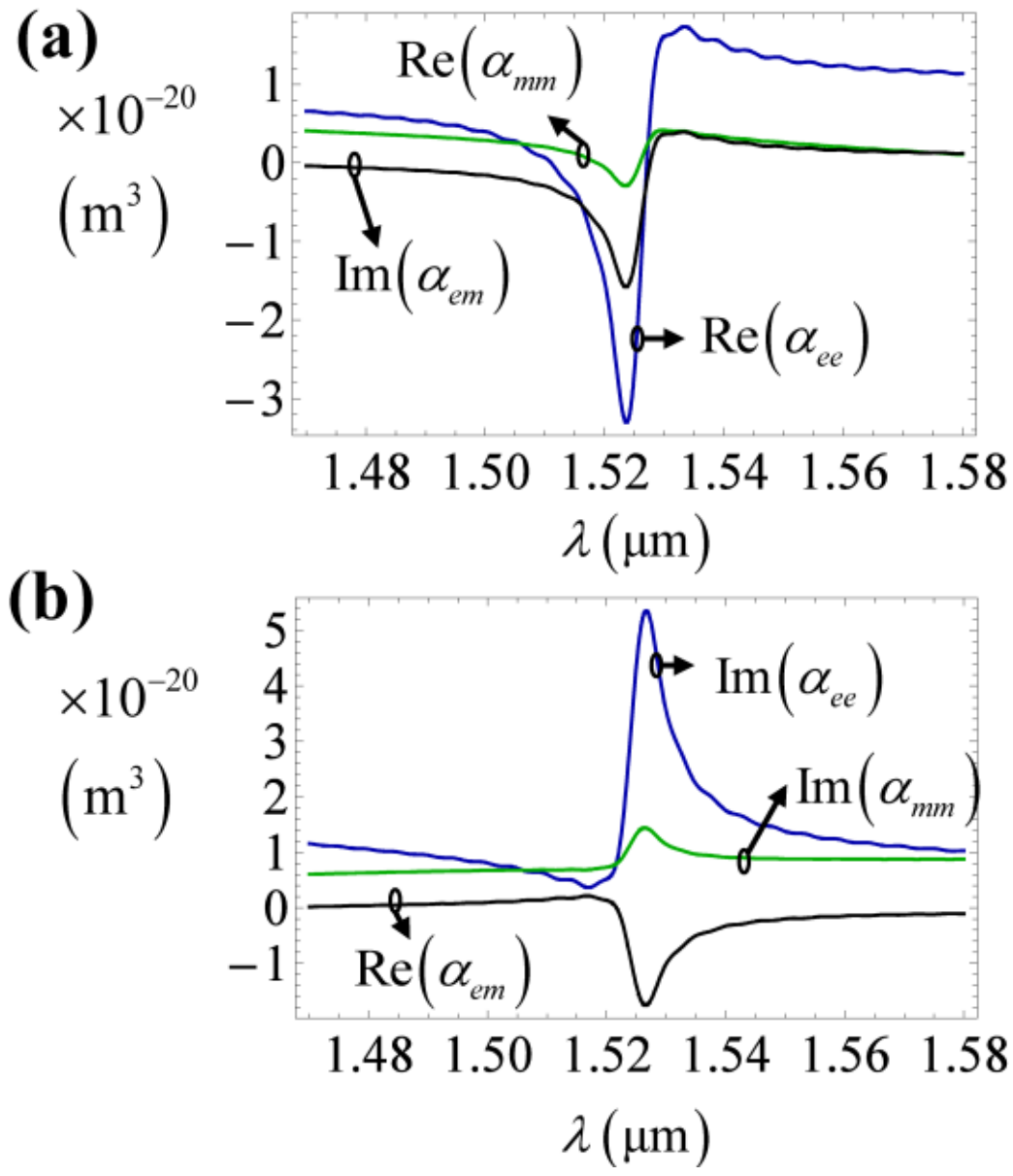

Fig. 4 Real and imaginary parts of the polarizabilities $\alpha_{ee}$ (blue– dark gray curves), $\alpha_{em}$ (black curves), and $\alpha_{mm}$ (green – light gray curve) of the chiral particle as a function of the wavelength.

Specifically, the chiral cuboid is illuminated by plane waves with the electric field oriented either along the $x$-direction ($\mathbf{E}^{inc} = E_0\hat{\mathbf{x}}$) or along the $y$-direction ($\mathbf{E}^{inc} = E_0\hat{\mathbf{y}}$), and that propagate either along the $+z$-direction or along the $-z$-direction, corresponding to a total of four different scattering problems. For each scattering problem, we determine the scattered far-field along the $\pm z$-directions, and finally the particle polarizabilities are written in terms of these scattered fields [49]. The scattered far-fields were obtained using the commercial full wave electromagnetic simulator CST Microwave Studio [51]. The numerically calculated polarizabilities of the chiral cuboid are represented in Fig. 4. The computed polarizabilities correspond to $\alpha_{ij} \equiv \hat{\mathbf{x}} \cdot \overline{\overline{\alpha}}_{ij} \cdot \hat{\mathbf{x}} = \hat{\mathbf{y}} \cdot \overline{\overline{\alpha}}_{ij} \cdot \hat{\mathbf{y}}$ with *i,j=e,m*. The polarizabilities of the chiral particle are resonant near $\lambda_0 = 1.53\,\mu\text{m}$ and the electric resonance ($\alpha_{ee}$) dominates. Crucially, the magneto-electric response is non-zero and $\alpha_{em} = -\alpha_{me}$ is the second strongest polarizability. In particular, $\text{Re}\{\alpha_{em}\}$ (related to the material loss) is peaked near $\lambda_0 = 1.53\,\mu\text{m}$ suggesting a strong discrimination between the optical forces induced by RCP and LCP incident waves ($\mathscr{F}_{z,RCP} - \mathscr{F}_{z,LCP} \approx 8k_0 W_{\text{av}}^{inc} \text{Re}\{\alpha_{em}\}$). To highlight the strongly asymmetric response of the chiral particle under LCP and RCP plane wave incidence, we calculated the spatial distribution of the electric field at the resonance (Fig. 5).

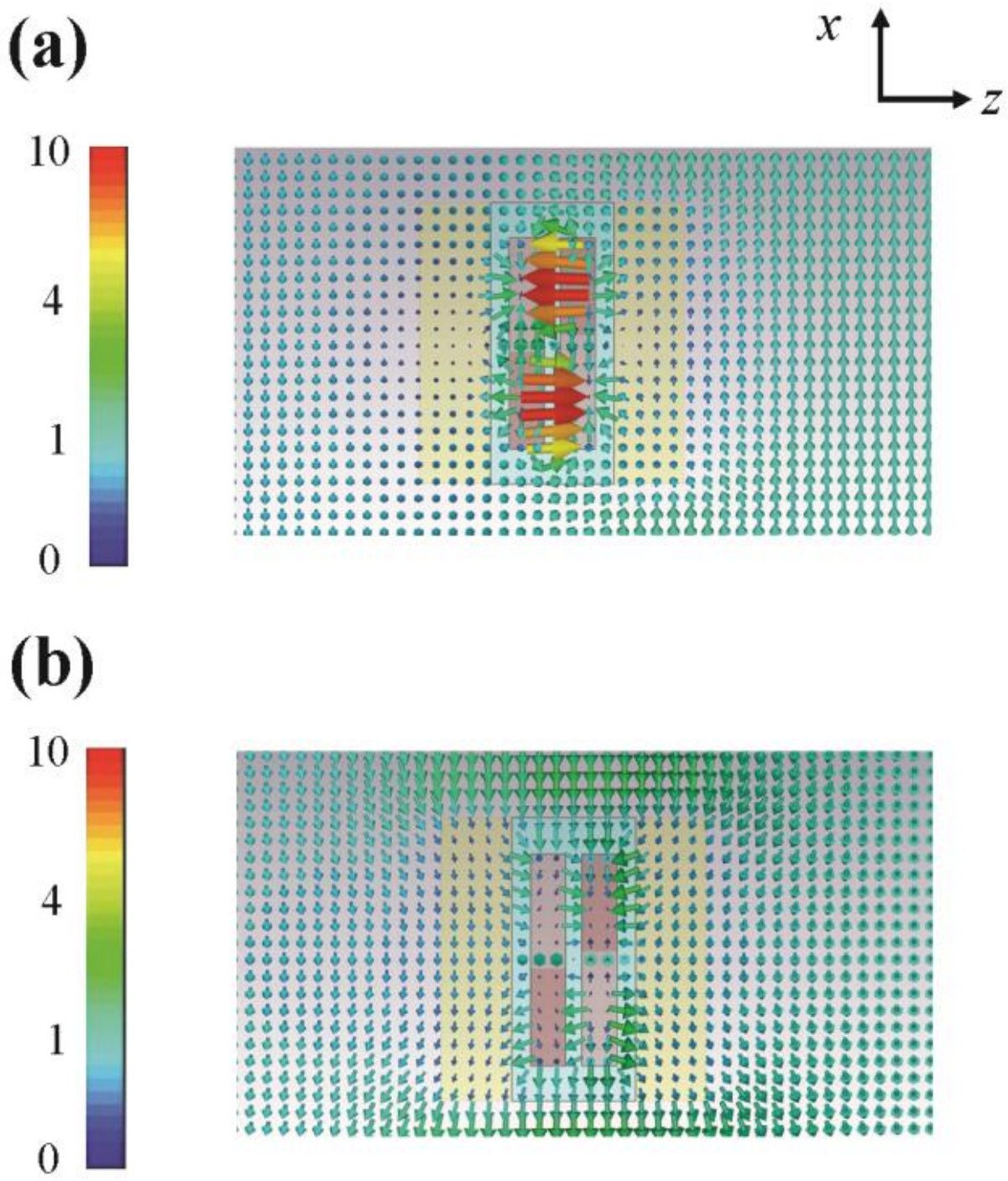

Fig. 5 **(a)** Density plot of the electric field (in arbitrary units) when the incident wave is RCP. **(b)** Similar to **(a)** but for an incident LCP wave.

As seen in Fig. 5, when the incident wave is LCP the incident field interacts weakly with the cuboid particle, whereas for an RCP excitation there is a resonant interaction leading to a strong absorption of light by the metamaterial particle. This confirms that the conjugated-gammadion particle interacts, indeed, very differently with RCP and LCP waves.

Using the formalism of Sect. II we calculated the optical force acting on the cuboid chiral particle when it stands alone in free space. We use the approximation that the electromagnetic response is isotropic, which as previously mentioned is expected to be very satisfactory because the *x*- and *y*- components of the fields are dominant. Figure 6 shows the normalized optical force ( $\mathcal{F}_z / a^2 W_{\text{av}}^{inc}$ ) exerted by RCP and LCP downstream excitations as a function of the particle position for $\lambda_0 = 1.53\mu\text{m}$. Note that by symmetry the lateral force must vanish and hence the only non-trivial component of the force is directed along *z*.

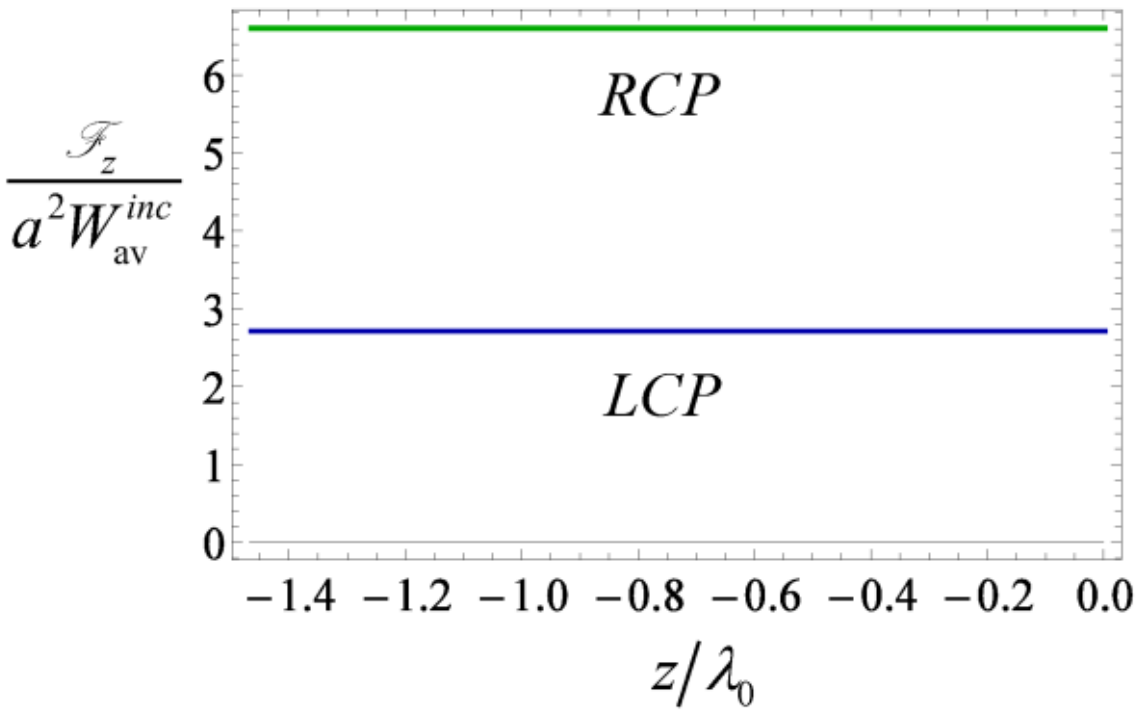

Fig. 6 Normalized optical force $\mathcal{F}_z / a^2 W_{\text{av}}^{inc}$ as a function of the particle position when the chiral particle stands alone in free space at $\lambda_0 = 1.53\mu\text{m}$. Blue curves: LCP incident light; Green curves: RCP incident light.

The results depicted in Fig. 6 confirm that the chiral response induces asymmetric optomechanical interactions. In the present scenario, the optical force is constant and positive, so that the particle is always pushed towards the downstream direction. Moreover, the optical force is about 2.5 times larger for RCP light because of the stronger light absorption for this polarization.

As outlined in Sect. II, one can take advantage of the peculiar electromagnetic response of the chiral cuboid to realize a passive optical conveyor belt. This is illustrated in Fig. 7a, which represents the optical force calculated when the particle is placed in front of the opaque mirror. As seen, in agreement with the general findings of Sect. II.D, because the response of the chiral cuboid is unbalanced ($\mathrm{Re}\left\{\alpha_{ee}-\alpha_{mm}\right\}\neq 0$) the force varies with the particle position. We only show the force up to a distance equal to $l=a_z$ because the particle has a certain thickness ($a_z\approx 0.124\lambda_0$) and the dipolar model is meaningless for shorter distances. Indeed, for shorter distances the effect of higher order multipoles needs to be taken into account, and a more sophisticated calculation method is required. In the range $l>a_z$, we find that the results obtained with Eqs. (18) (dashed curves in Fig. 7a) follow very closely the value of the force calculated with the exact formula (solid lines). In our calculations it is assumed that the orientation of the particle with respect to the incoming wave is fixed. In general, the macroscopic optical anisotropy or even the Brownian dynamics may influence the stability of the particle orientation. This issue can be avoided with the proper engineering of the chiral particle, for example by considering a truly isotropic (omnidirectional) chiral design [28, 50].

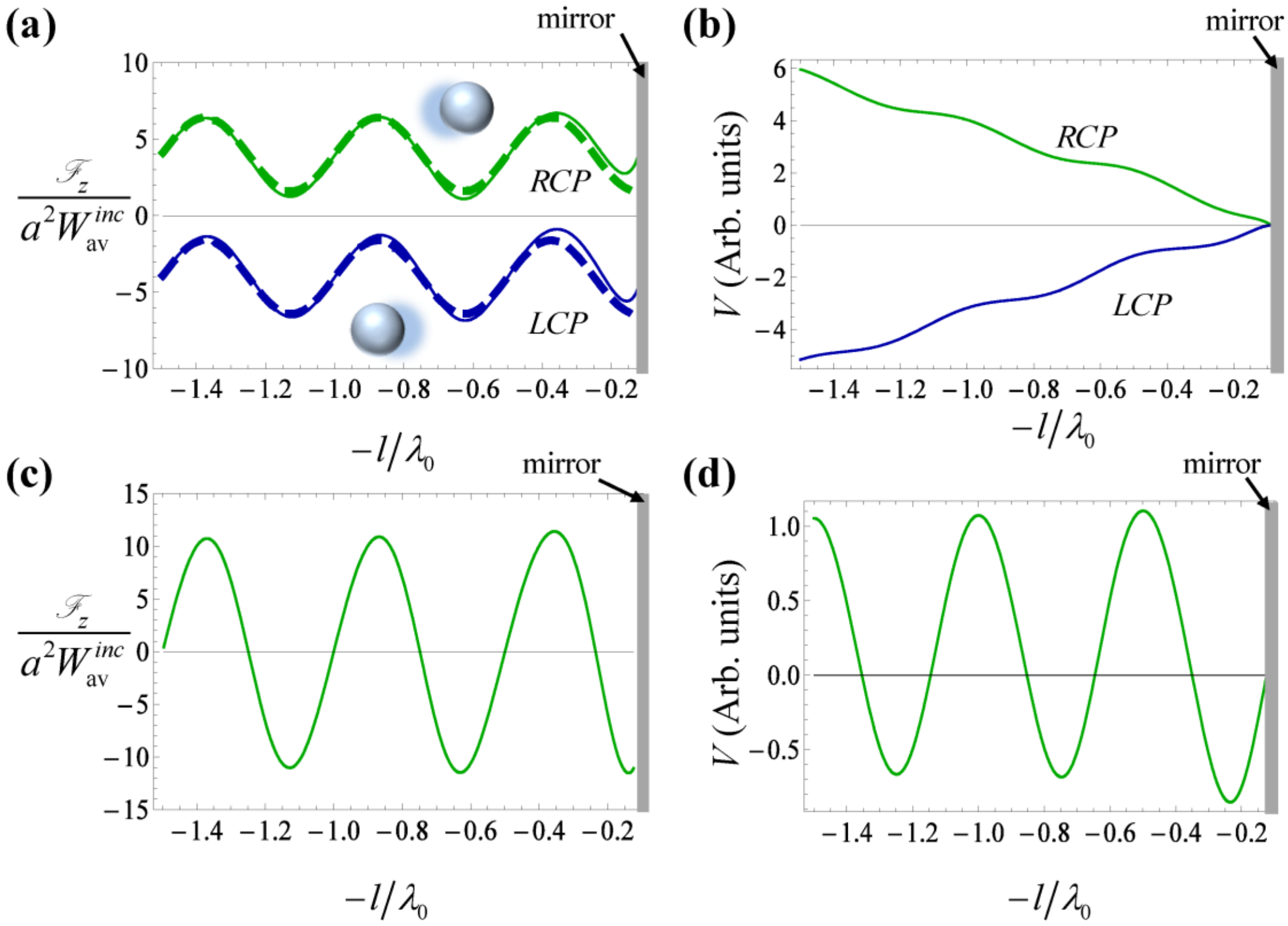

Fig. 7 **(a)** $\mathcal{F}_z$ as a function of the distance $l$ to the mirror at $\lambda_0 = 1.53\mu\text{m}$ for LCP incident light (blue curves) and RCP downstream light (green curves) calculated with the exact formalism (solid curves) and neglecting the contribution from the image dipoles (dashed curves). **(b)** Potential (in arbitrary units) as a function of the distance $l$. **(c)** and **(d)** Similar to **(a)** and **(b)** respectively, but for a dielectric (polyimide) sphere with radius $r = a_z$.

Importantly, the sign of the optical force for a fixed helicity of the incoming wave is always the same. Indeed, it can be checked that the polarizabilities of the chiral particle satisfy the constraint (19). In particular, for RCP incidence the optical force is always positive so that the particle is steadily pushed towards the mirror, whereas for LCP incidence the force is repulsive so that the particle is steadily pulled towards the light source. Thus, the proposed configuration allows for an optical conveyor belt operation without optical traps. In agreement with this property, the potential energy $V = -\int \mathcal{F}_z dz$ varies monotonically with the distance to the mirror for both polarizations (Fig. 7b). Thus, by controlling the intensity and polarization of the incident wave it may be possible to judiciously control the position of the particle. High-power lasers under continuous wave operation may generate a light intensity $S^{inc} \approx 20\,\mathrm{GW/m^2}$ [52] at $\lambda_0 = 1.55\,\mu\mathrm{m}$, and hence the peak optical force can be on the order $\mathcal{F}_z \approx 15\mathrm{pN}$, which has a magnitude estimated to be about $5.6 \times 10^4$ times larger than the gravity force acting on the chiral cuboid. The optical force usually obtained with optical tweezers spans from subpiconewtons to 100 pN [7].

To highlight the unique optical manipulations enabled by chiral particles, we contrast the response of the gammadion particle with that of an electrically small dielectric sphere. We consider a sphere with radius $r = a_z$ made of polyimide. The electromagnetic response of a small dielectric sphere is completely characterized by the electric polarizability $\alpha_e = 1/\left(\alpha_0^{-1} - i\,k_0^3/6\pi\right)$, being $\alpha_0 = 4\pi r^3 \dfrac{\varepsilon_h - 1}{\varepsilon_h + 2}$ the quasi-static polarizability given by the Clausius-Mossotti relation [45, 53]. The calculated optical force acting on the dielectric sphere at $\lambda_0 = 1.53\mu\mathrm{m}$ is depicted in Fig. 7c). The results reveal that the sign of the force depends on the distance of the nanoparticle with respect to the mirror, which is markedly different from the optical conveyor belt regime. This behavior can also be understood from Eq. (18). Indeed, in the absence of a magneto-electric response, the optical force is the same for both polarizations and has a sinusoidal-type spatial variation of the form $\sin\left(2k_0 z_0\right)$. Importantly, the co-existence of repulsive and attraction regions leads to the formation of potential wells wherein the dielectric sphere can be trapped. This feature is demonstrated in Fig. 7d) where we show the potential energy as a function of the distance to the mirror. This regime is analogous to the standard optical traps [54] wherein particles can be trapped in the locations wherein the optical field has an intensity maximum.

# IV. Conclusion

We theoretically proposed a different approach to transport tailored metamaterial nanoparticles using a single beam optical conveyor belt. It was demonstrated that by controlling the helicity of the incoming wave it is possible to switch between persistent attractive or repulsive optical forces, with signs independent of the particle position. In the ideal case the nanoparticle should be balanced, such that the real parts of the electric and magnetic polarizabilities are identical. In such a situation the gradient force vanishes and the intensity and the sign of the optical force are independent of the particle position with respect to the mirror. We derived a simple condition that ensures a passive optical conveyor belt operation for unbalanced particles [Eq. (19)]. The theory was numerically illustrated by considering a tailored metamaterial cuboid formed by two conjugated gammadions. Even though our design is not optimized and the particle response is not balanced, the simulations indicate that there is a strong discrimination between the optical forces acting on the metamaterial particle for incoming waves with different helicities. Remarkably, for realistic values of the incoming light intensity the peak optical force exceeds the gravity force by about four orders of magnitude in the regime where the force is negative. Hence, we believe that an experimental verification of our proposal may be within reach.

**Acknowledgements**

This work was partially funded by Fundação para Ciência e a Tecnologia under project PTDC/EEI-TEL/4543/2014.